\documentclass[aps,prl,twocolumn,showpacs,preprintnumbers,amsmath,amssymb]{revtex4}
\usepackage{amsmath}
\usepackage{graphicx}
\usepackage{psfrag}
\usepackage{natbib}
\begin{document}
\title{Quantum Dynamics of Electron-Nuclei Coupled System in
Quantum Dots }

\author{ \"Ozg\"ur \c{C}ak{\i}r and Toshihide Takagahara}

\affiliation{Department of Electronics and Information Science,
Kyoto Institute of Technology,
Matsugasaki, Kyoto 606-8585 JAPAN}

\affiliation{ \\}

\affiliation{CREST, Japan Science and Technology Agency, 4-1-8 Honcho,
Kawaguchi,
Saitama 332-0012, JAPAN}

\date{\today}

\begin{abstract}

We have investigated the dynamics of the electron-nuclei
coupled system in quantum dots. The bunching of
results of the electron spin measurements and the revival in the conditional
probabilities are salient features of the nuclear spin memory. The underlying
 mechanism is the squeezing of
the nuclear spin state and the correlations  between the successive electron spin measurements.
%This squeezing
%is expected to lead to the elongation of the electron spin coherence time.
Further we make a proposal for the preparation and detection of superposition states of nuclear spins
merely relying on electron spin measurements. For unpolarized, completely random nuclear spin state one can still trace the quantum interference effects. We discuss the realization of these schemes for electron spins on both single and double QDs.
\end{abstract}
\pacs{73.21.La, 71.70.Jp, 76.70.-r, 03.67.Pp}
\maketitle

Electron spins in semiconductor quantum dots(QDs) are considered as one of the
most promising candidates of the building blocks for quantum information
processing\cite{Loss98,Imamoglu99}  due to their robustness
against decoherence effects\cite{Golovach04,Semenov04}.
In double QD systems, initialization and coherent manipulation of electron spin have been realized, with coherence times extending to $1~\mu{\tt s}$\cite{Petta05, Koppens06}.
Hyperfine(HF) interaction with the host nuclei\cite{Abragam96, Merkulov02} becomes the main decoherence mechanism, dominating over
spin-orbit interactions which act on a timescale of 10s of milliseconds\cite{Kroutvar04,Meunier06} or even longer. Consequently there have been proposals to reduce HF induced decoherence by measuring or polarizing the nuclear
spins\cite{Taylor03b,Imamoglu03,Klauser05,Giedke05,Stepanenko06,Rudner06} and to use nuclear spins as a quantum memory\cite{Taylor03a,Taylor03b}.

Here we investigate the electron-nuclei spin coupling in QDs, and show that consecutive electron spin measurements following
 HF interaction are correlated and lead to purification of the nuclear spin system.
 %More specifically, starting from an unknown initial state of nuclear
%spins, successive measurements of the electron spin states result in narrowing of the
%distribution of the nuclear spin field.
We predict that the purification of the nuclear spin state
 would lead to the bunching of results of the electron spin state
measurements and also to the reduction in the electron
spin decoherence induced by the HF interaction. We will also discuss a strategy for revealing quantum nature of nuclear spins from the correlations of successive electron spin measurements.
For the  physical realization of the proposals we will in particular discuss a double QD occupied by two electrons, and a single QD occupied by one(two) electron(s).

First of all we consider an electrically gated double QD occupied by two electrons\cite{Petta05,Coish05}.
The excited electronic orbitals of QDs have an energy much greater than the thermal energy and the adiabatic voltage sweeping rates, so that the electrons occupy only the ground state orbitals.
Under a high magnetic field, s.t. the electron Zeeman splitting is much greater than
 the HF fields and the exchange energy, dynamics takes place in the spin singlet ground state $|S\rangle$ and triplet state of zero magnetic quantum number $|T\rangle$. For the singlet state each electron can be found in the different or both in the same QD, whereas for the triplet state electrons can only be found in different QDs.
 %For brevity the spin notation $|S(T)\rangle$ also involves the orbital wavefunctions.
 Singlet and triplet states are coupled by the HF fields, and the system is governed by the Hamiltonian,
 \begin{eqnarray}
H_e=JS_z+ r\delta h_z S_x, \label{eq_hf}
\end{eqnarray}
%The electrons also feel the HF magnetic fields of the host nuclei\cite{Abragam96}.  In the $(1,1)$ charge configuration,  the ground state orbitals are well localized so that HF fields in different dots can be regarded as independent. On the other hand, when spin singlet electrons are in the same QD, viz. $|02;S\rangle$ state, HF interaction vanishes.
 where ${\bf S}$ is the pseudospin operator with  $|T\rangle$ and $|S\rangle$ forming the $S_z$ basis.
  $J$ is the exchange energy and
 $\delta h_z=h_{1z}-h_{2z}$, where $h_{1z}$ and $h_{2z}$ are the components of nuclear HF field along the external magnetic field in the first and second dot, respectively.
 $0\le r\le 1$ is the amplitude of the hyperfine coupling.  When both electrons are localized in the same dot, $r\rightarrow 0$ and $J\gg \delta h_z$, when they are located in different dots HF coupling is maximized $r\rightarrow 1$ and $J\rightarrow 0$.

 Now we show that by electron spin measurements in a double QD governed by (\ref{eq_hf}), the coherent behavior of nuclear spins can be demonstrated.
 Electron spins are initialized in the singlet state and the nuclear spin states are initially in a mixture of  $\delta h_z$ eigenstates,
$\rho(t=0)=\sum_n p_n\rho_n|S\rangle\langle S|$,
where $\rho_n$ is a nuclear state with an eigenvalue of $\delta h_z=h_n$ and satisfies $Tr(\rho_n)=1$. $p_n$ is the probability of the hyperfine field $\delta h_z$ having the value $h_n$.
In the unbiased regime $r=1$, the nuclear spins and the electron spins interact for a time span of $\tau$.
%leading to the state, $
%\rho=\sum_n p_n\rho_n|\Psi_n\rangle\langle \Psi_n|$,
% where $|\Psi_n\rangle=\alpha_n|S\rangle+\beta_n|T\rangle$, with
% $\alpha_n=\cos\Omega_n \tau/2+iJ/\Omega_n \sin\Omega_n \tau/ 2$,
% $\beta_n=
% -ih_n/\Omega_n\sin\Omega_n \tau/2$
 %and  $\Omega_n=\sqrt{J^2+h_n^2}$.
  Then
the gate voltage is swept adiabatically, switching off the HF interaction $r\rightarrow 0$, in a time scale much shorter than HF interaction time.
Next a charge state measurement is performed which detects a singlet or triplet state. Probability to detect the singlet state is $\sum_n p_n|\alpha_n|^2$, and the triplet state is $\sum_n p_n|\beta_n|^2$ where
$\alpha_n=\cos\Omega_n \tau/2+iJ/\Omega_n \sin\Omega_n \tau/ 2$, $\beta_n=
 -ih_n/\Omega_n\sin\Omega_n \tau/2$, with $\Omega_n=\sqrt{J^2+h_n^2}$.
Subsequently one can again initialize the system in the singlet state of electron spins, and  turn on the hyperfine interaction for a time span of $\tau$, and perform a second measurement.
In general over $N$ measurements, the nuclear state conditioned on $k(\leq N)$ times singlet and $N-k$ times triplet detection is $
\sigma_{N,k}=\bigl(^N_{\,k}\bigr)\sum_n p_n|\alpha_n|^{2k}|\beta_n|^{2(N-k)}\rho_n,\label{nucstate}$
the trace of which yields the probability of $k$ times singlet outcomes,
 \begin{align}
 P_{N,k}=Tr\sigma_{N,k}
 =\bigl(^N_{\,k}\bigr)\langle|\alpha|^{2k}|\beta|^{2(N-k)}\rangle,\label{Pqm}
 \end{align}
 where  $\langle\ldots\rangle$ is the ensemble averaging over the hyperfine field $h_n$\cite{Merkulov02}.
 Hereafter, this case will be referred to as the {\it coherent regime}.
%Here the key assumption is that nuclear states preserve their coherence over $N$ measurements, thus the measurements are not independent due to nuclear memory.
One can easily contrast this result with that for the {\it incoherent} regime in which nuclear spins  lose their coherence in between the successive spin measurements and relax to the equilibrium distribution, given by
%In the {\it incoherent regime} of nuclear spins, results of successive measurements
% are independent and the probability for obtaining $k$ times singlet results over $N$ measurements is given by,
\begin{eqnarray}
 P'_{N,k}=\bigl(^N_{\,k}\bigr)\langle|\alpha|^2\rangle^{k}\langle|\beta|^2\rangle^{(N-k)}.\label{Psc}
\end{eqnarray}
%When the nuclear spins are incoherent, the probability distribution (\ref{Psc}) obeys simply a Gaussian distribution with mean $k=N\langle|\alpha|^2\rangle$, and variance $N\langle|\alpha|^2\rangle\langle|\beta|^2\rangle$, as $N\rightarrow\infty$.
% However, when nuclear spins preserve their coherence, the probability distribution (\ref{Pqm}) may exhibit different statistics depending on the initial nuclear state.
% The two probability distributions (\ref{Pqm}) and (\ref{Psc}) yield the same mean value, $
%\overline{k}=N\langle|\alpha|^2\rangle$,
 %however with distinct higher order moments. If the equlibrium distribution of nuclear state $p_n$ has a width $\Delta$, then for HF interaction time $\tau\geq 1/\Delta$, the distributions (\ref{Pqm},\ref{Psc}) start to deviate from each other.
 %They yield the same distribution only when the initial nuclear state is in a well defined eigenstate of $\delta h_z$, i.e. when $\Delta=0$.

%As the simplest case let us check the results of two measurements, $N=2$. The probability for two singlet
%detections are given by
%$P_{2,2}=\langle|\alpha|^4\rangle$,
%$ P'_{2,2}=\langle|\alpha|^2\rangle^2$  and  $P_{2,2}\geq P'_{2,2}$.
%However,
%for one singlet detection, $P_{2,1}=2\langle|\alpha|^2-|\alpha|^4\rangle$,
%$ P'_{2,1}=2(\langle|\alpha|^2\rangle-\langle|\alpha|^2\rangle^2)$, respectively and this time $P'_{2,1}\geq P_{2,1}$.
If the nuclear spins are coherent over the span of the experiment, then
successive electron spin measurements are biased to all singlet(triplet) outcomes.
In particular, when the initial nuclear spins are unpolarized and randomly oriented, the distribution of hyperfine field is characterized by a Gaussian distribution with variance $\sigma^2$,
$p[h]=\frac{1}{\sqrt{2\pi\sigma^2}}e^{-\frac{h^2}{2\sigma^2}}$
and the summation is converted to an integration, $\sum_np_n\ldots\rightarrow \int {\tt d}h\,p[h]$.
As the simplest case, let us check the results of two measurements, each following a HF interaction of duration $t$. The probability for two singlet detections are given by
$P_{2,2}=\langle|\alpha|^4\rangle=\{6+2 e^{-2t^2}+8e^{-t^2/2}\}/16$, which is always greater than $ P'_{2,2}=\langle|\alpha|^2\rangle^2=\{4+8e^{-t^2/2}+4e^{-t^2}\}/16$, results given particularly for $J=0$.
As $J$ is increased the probabilities approach each other and for $J\gg \sigma$ they become identical\cite{Cakir06}.

%In Fig. \ref{Fig_2meas}, $P_{2,0}$, $P'_{2,0}$ and $P_{2,1},P'_{2,1}$ are shown as a function of HF interaction time $\tau$, for two cases of the exchange energy $J=0$ and $J=0.5\sigma$. Two spin singlet, or triplet measurement probability is enhanced, while one singlet-one triplet measurement probability is depressed in coherent picture of nuclear spins.
%As the exchange energy is increased, singlet-triplet mixing is suppressed, thus coherent and incoherent pictures approach each other.

%\begin{figure}[h!]
%\includegraphics[width=8.6cm]{fig_1.eps}
%\caption{ Probability of outcomes of two measurements  as a function of HF interaction time $\tau$ : One singlet-one triplet(coherent regime, solid line; incoherent regime, dashed line), two triplets(coherent regime, dotted; incoherent regime, dash-dotted curve)  for   a) $J=0$, b) $J/\sigma=0.5$ .\label{Fig_2meas}}
%\end{figure}
\begin{figure}[!t]
\includegraphics[width=8.6cm]{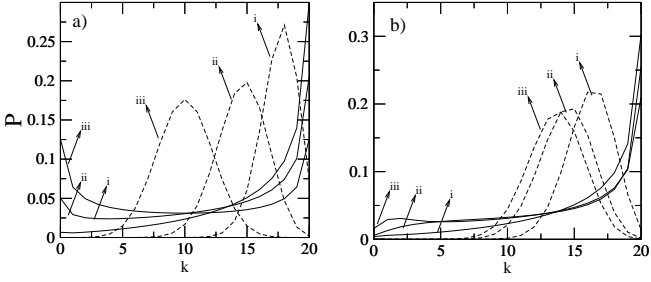}
\caption{Probability distribution  at $N=20$ measurements for $k={0,1,\ldots,20}$ times singlet detections, for coherent regime(solid lines) and incoherent regime(dashed lines). Two cases of the exchange energy are considered  a) $J=0$ b) $J/\sigma=0.5$ for HF interaction times $\sigma\tau=$ i)$0.5$, ii)$1.5$, iii)$\infty$. \label{Fig_20meas}}
\end{figure}

In Fig. \ref{Fig_20meas}, for $N=20$ measurements, $P_{N,k}$ is shown
 for HF interaction times $\sigma\tau=0.5, 1.5, \infty$. For $\tau=0$, the probability for both (\ref{Pqm}) and (\ref{Psc}) is peaked at
$k=20$. However, immediately after the HF interaction is introduced, the probability distributions show distinct behavior.  The measurement results in the incoherent regime approach a Gaussian distribution. In the coherent case the probabilities
bunch at $k=0,20$ for $J=0$, and when $J/\sigma=0.5$ those bunch at $k=20$ only. As $J$ is increased above some critical value, no bunching takes place at $k=0$ singlet measurement.

%To observe the bunching $N$ succesive spin measurements are performed within the coherence time of the nuclear spins, then after waiting for some time so that nuclear spins are again randomized, another set of
%$N$ successive measurements are carried out and so on. Thus an ensemble averaging of $N$ measurements is performed which results in a bunching of either spin singlet or triplet outcomes. This bunching is a clear sign of coherent behavior of nuclear spins, which can easily be contrasted with the incoherent regime which merely exhibits a Gaussian distribution.

 %Lacking any prior knowledge about the nuclear states one can gather some information about the initial nuclear state by conditional measurements. This information can be used to decrease the effect of HF interaction.
%After performing $N$ measurements all with the same HF interaction time $\tau$, conditioned on the detection of $k$ times singlet and $N-k$ times triplet states, the nuclear state is given by Eq. (\ref{nucstate}) with an appropriate normalization constant.

\begin{figure}[!t]
\includegraphics[width=8.6cm]{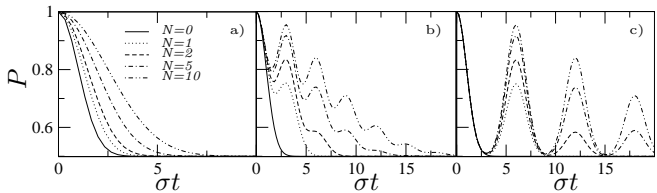}
\caption{Conditional probability for singlet state detection as a function of HF interaction time $\sigma t$, subject to $N=0,1,2,5,10$ times prior singlet state measurements and for HF interaction times a)$\sigma\tau=1.0$, b)$\sigma\tau=3.0$, c)$\sigma\tau=6.0$. \label{Fig-cond}}
\end{figure}

The nuclear spin state conditioned on the previous electron spin measurements is no longer random even if they are initially random. Accordingly, HF induced electron spin decoherence dynamics is also modified.
Depending on the results of previous measurement, one may decrease the singlet-triplet mixing.
% For instance, detection of singlet states will lead to modulation of nuclear spectrum $p[h]\rightarrow p[h]\cos^2h\tau/2$.
As a particular example, consider the case: Starting from a random spin configuration, $N$ successive
electron spin measurements are performed, each following initialization of electron spins in the spin singlet state and a HF interaction of duration $\tau_i$($i=1\ldots N$) and all outcomes turn out to be singlet. Here the nuclear spin state is given by  $\sigma_{N,N}$. Then again HF interaction is switched on for a time $t$, and the $(N+1)$th measurement is carried out. The conditional probability to detect the singlet state is given by
\begin{eqnarray}
 P=\frac{\sum(^{~2}_{s_1})(^{~2}_{s_2})\ldots(^{~2}_{s_{N+1}})e^{-\frac{1}{2}\bigl[\sum_{i=1}^N(s_i-1)\tilde{\tau}_i+(s_{N+1}-1)\tilde{t}\bigr]^2}}{
4\sum(^{~2}_{s_1})(^{~2}_{s_2})\ldots(^{~2}_{s_N})e^{-\frac{1}{2}\bigl[\sum_{i=1}^N(s_i-1)\tilde{\tau}_i\bigr]^2} },
 \label{condprob}
 \end{eqnarray}
 where the sums run over $s_i=0,1,2$ and $\tilde{\tau}_i=\sigma\tau_i$.
 For the particular case $\tau_1=\tau_2=\ldots=\tau_N=\tau\gg
1/\sigma$,
 the initial state is revived at $t=n\tau, \,(n=1,2,\ldots,N)$ with
 a decreasing amplitude, $P\simeq1/2+\sum_{s=0}^{N}(
^{2N}_{~s})e^{\frac{-\sigma^2}{2}(t-(N-s)\tau)^2}/4(^{2N}_{~N})$.
 In Fig. \ref{Fig-cond} the conditional probabilities(\ref{condprob}) are shown for $\sigma\tau=1.0, 3.0, 6.0$ subject to $N=0,1,2,5,10$ times prior singlet measurements in each. Revivals are observable only for $\sigma\tau>1$, because the modulation period of the nuclear state spectrum characterized by $1/\tau$ should be smaller than the variance $\sigma$.
 The underlying mechanism of revivals is purification of nuclear
spins by the electron spin measurements. The purity of a system
characterized by the density matrix $\hat{\rho}$ is given by ${\cal
P}=Tr\hat{\rho}^2$. As an example we are again going to consider the
nuclear state prepared by $N$ successive electron spin measurements
with singlet outcomes, each following a HF interaction of duration 
$\tau_{1}\ldots \tau_N$. The purity of nuclear spins is given by
\begin{eqnarray}
%{\cal P}&=\frac{\int{\tt
%d}h~p[h]\cos^4\frac{h\tau_1}{2}\cos^4\frac{h\tau_2}{2}\ldots\cos^4\frac{h\tau_N}{2}}{\int{\tt
%d}h~p[h]\cos^2\frac{h\tau_1}{2}\cos^2\frac{h\tau_2}{2}\ldots\cos^2\frac{h\tau_N}{2}}\nonumber\\
{\cal P}=\frac{1}{\cal D}\frac{\sum_{s_i=0}^
4(^{~4}_{s_1})(^{~4}_{s_2})\ldots(^{~4}_{s_N})e^{-\frac{1}{2}\bigl[\sum_{i=1}^N(s_i-2)\tilde{\tau}_i\bigr]^2}}{\bigl[ \sum_{s_i=0}^
2(^{~2}_{s_1})(^{~2}_{s_2})\ldots(^{~2}_{s_N})e^{-\frac{1}{2}\bigl[\sum_{i=1}^N(s_i-1)\tilde{\tau}_i\bigr]^2}\bigr]^2 },\label{purity}
\end{eqnarray}
where ${\cal D}$ is the dimension of the Hilbert space for the nuclear spins. For a fixed ratio of $\tau_1:\tau_2:\ldots:\tau_N$, purity
(\ref{purity}) is a monotonically increasing function of time. For
$\sigma\tau_i\gg 1$, one can attain various asymptotic limits for
the purity.  For instance, for $N=2$, there are three asymptotic
limits;when a)$\tau_1=2\tau_2$ then ${\cal P}=11/4{\cal D}$,
b)$\tau_1=\tau_2$ then ${\cal P}=35/18{\cal D}$, c)otherwise ${\cal P}=9/4{\cal D}$.
For $N=2$ with $\tau_1=2\tau_2=2\tau\gg 1/\sigma$,  the conditional
probability (\ref{condprob}) is given as,
$P\simeq 1/2+ \sum_{n=0}^3(4-n)\exp[-(\tilde{t}-n\tilde{\tau})^2/2 ]/8$
whereas for $\tau_2=\tau_1=\tau\gg 1/\sigma$,
$P\simeq 1/2+\bigl\{ e^{-\frac{(\tilde{t}-2\tilde{\tau})^2}{2} }+
4e^{-\frac{(\tilde{t}-\tilde{\tau})^2}{2} }+6e^{-\frac{\tilde{t}^2}{2} }
\bigr\}/12$.
It can be seen that as the purity of nuclear spins increases, more revivals are present
with an increased amplitude.

So far we have discussed the bunching and revival phenomena only for a double QD system. The same predictions can also be  made for a single QD occupied by a single electron\cite{Hanson03,Dutt05,Atature06}.
Consider a single QD occupied by a single electron, under an external magnetic field s.t. electron Zeeman energy is much greater than the HF energies. Then the system is described by the Hamiltonian, $H\simeq
g_e\mu_BB S_z + h_z S_z$. Here $g_e$ is the electron $g$-factor, $\mu_B$ the Bohr magneton and $B$ is the external field applied in $\hat{z}$ direction. Spin flips are suppressed since $g_e\mu_B B\gg \sqrt{\langle {\bf h}^2\rangle}$. $|\pm\rangle=(|\uparrow\rangle\pm|\downarrow\rangle)/\sqrt{2}$ states
are coupled by HF interaction  with $|\uparrow(\downarrow)\rangle$
being the eigenstates of $S_z$. Each time the electron is prepared
in $|+\rangle$. Next it is loaded onto the QD, then removed from the
QD after some dwelling time $\tau$. Next spin measurement is
performed in $|\pm\rangle$ basis.
 Essentially the same predictions
as those for double QD can be made for this system, namely electron
spin bunching and revival.
%In Fig. \ref{figSQD}, for $N=40$ measurements, the QM probability
%distribution of $P_{N,k}$ is shown at electron Zeeman energy
%$\epsilon=g_e\mu_B B=3\sigma$, for $\sigma\tau=0.3,~,0.6,~0.9,~\infty$. It is
%seen, due to external field the population bunches at
%$|-\rangle$ states at times $\tau\sim \pi/\epsilon$, but then relaxes to the
%equilibrium distribution for $\sigma\tau\gg 1$ which is the same equilibrium distribution as that of double QD.
%Next we are going to consider electron spin revivals.
We are going to consider electron spin revivals as an example.
After $N$ times HF interaction of duration $\tau\gg 1/\sigma$, each
followed by $|+\rangle$ measurement, the conditional probability for
obtaining $|+\rangle$ in the $(N+1)$th step following a HF
interaction of duration $t$ is given as, $ P\simeq
1/2+\sum_{s=0}^{N}(
^{2N}_{~s})e^{-\sigma^2(t-(N-s)\tau)^2/2}\cos\epsilon[t-(N-s)\tau]/4(^{2N}_{~N})$.
%\begin{figure}
%\psfrag{a}{$k$} \psfrag{b}{$P$}
%\includegraphics[width=18pc]{figSQDbunching}
%\caption{Single QD: Probability distribution  at $N=40$ measurements
%for $k={0,1,\ldots,40}$ times $|+\rangle$ detections at HF
%interaction times  $\sigma\tau=$ i)$0.3$, ii)$0.6$, iii)$0.9$
%iv)$\infty$. \label{figSQD}}
%\end{figure}

The Hamiltonian (\ref{eq_hf}) can also be used to describe a pair of electrons in a single QD\cite{Fujisawa02,Hanson05}, and the same predictions as those for a double QD can be made. In the two electron regime, the energy splitting between singlet ground state and triplet excited state can be tuned by application of a magnetic field\cite{Hanson05,Meunier06b}.
Detuning is given by $\Delta E\simeq \omega^2/\Omega$ for $\Omega>>\omega$, $\Omega$ being the electron Larmor frequency, and $\omega$ is the frequency of the harmonic confinement in the lateral plane.
Under a high magnetic field, the triplet state of zero magnetic quantum number is coupled to singlet state via the HF field.
This HF field is estimated to be about $0.1~\mu$eV and is comparable to $\Delta E$($\simeq 0.2~$meV) for a GaAs QD with $10$nm thickness and $\omega=0.1~$meV under a transverse magnetic field of $20~$T.
%$h=Av_0\sum \phi_g({\bf r}_i) \phi^*_e({\bf r}_i)I^{(i)}_z/\sqrt{2}$. $\phi_{g(e)}$ is the ground(excited) state orbital in the QD, $A$ the HF
%coupling constant\cite{Paget77}, $v_0$ the unit volume of the cell and $I^{(i)}_z$ are the nuclear spins.
%For a GaAs QD, of thickness $10$ nm, $\omega=0.1$ meV and transverse magnetic field $B=20$T, $\Delta E\simeq 0.3\mu$eV and is of the same
%order of magnitude as the HF field $\sigma$.
The electrons' spin state can be initialized and measured with high fidelity by a spin selective coupling to leads,
relying on spin dependent tunnel rates\cite{Hanson05}.

%$h=Av_0\sum \phi_g({\bf r}_i) \phi^*_e({\bf r}_i)I^{(i)}_z/\sqrt{2}$. $\phi_{g(e)}$ are the ground and the first excited state orbitals in the quantum dot, $A$ is the HF
%coupling constant\cite{Paget77} which is $\sim 90\mu eV$ for GaAs, and $v_0$ is the unit volume of the cell.  ${\bf I}^{(i)}_z$ are the components of nuclear spins along the external field.
%Typically for a two dimensional QD, with harmonic confinement the HF field has an {\it rms} value
%$\langle h^2\rangle=3A^2s(s+1)v_0/16\sqrt{2}\pi d\sigma^2$ where $\sigma=\sqrt{\hbar/m\tilde{\omega}}$ is the Fock-Darwin radius and $d$ is the thickness of the QD. Here $\tilde{\omega}=\sqrt{\omega^2+\Omega^2/4}$, $\Omega$ being the Larmor frequency of the electron, and $\omega$ is the frequency of the lateral isotropic harmonic confinement.
%The detuning between the triplet and the singlet states is $E_T-E_S=\hbar(\tilde{\omega}-\Omega/2)$.
%Typically for a quantum dot of $50 \AA$ radius, and thickness $d=10\AA$, strength ofthe HF coupling will be $\Delta h\sim 4\mu{\tt eV}$.

In cases so far considered, the quantum nature of nuclear spins is not manifest because the same predictions can be made using semiclassical picture of nuclear spins.  In order to detect the quantum behavior of nuclear spins, one has to prepare the nuclear spins
in superposition states. For the models under consideration, this can be achieved via switching HF interaction for different components of HF field for successive measurements. This can be realized by changing the direction of the external field. This enables one to observe the interference effects, since different components of the HF field  do not commute.
As an example we consider the case of a pair of electrons on a single QD
with homogeneous HF coupling throughout the dot, i.e. ${\bf h}=a\sum {\bf I}^{(i)}$ with $a=A/N_n$, $N_n$ being the number of nuclear spins.  By applying a magnetic field in $\hat{n}$ direction, s.t., electron Zeeman energy is much greater than the HF fields, an effective HF coupling of the form
$V=({\bf h}\cdot\hat{n})(|S\rangle\langle\hat{n};T_0|+{\tt h.c.})$ can be obtained, with $\hat{n}$ being the quantization axis. 
 The electron is initialized in the singlet state where detection is performed in singlet-triplet basis.
We have to perform a series of measurements which involve external field applied in different directions. As a particular example we will consider the conditional evolution of the
nuclear system described by $U_c={\cal M}_zU_z(\tau_3){\cal M}_xU_x(\tau_2){\cal M}_zU_z(\tau_1)$, where $U_{\hat{n}}(\tau)$ is the unitary evolution of duration $\tau$ following an electron spin initialization in singlet state. ${\cal M}_{\hat{n}}$ is the electron spin measurement in $|S\rangle$,$|\hat{n};T_0\rangle$ basis.
  Initially starting from an ensemble of nuclear spins polarized in $\hat{z}$ direction,
\begin{eqnarray}
\rho(t=0)=\sum_{\lambda,j,m} \frac{P[m]}{F[m]}|\lambda j m\rangle\langle\lambda j m|
\end{eqnarray}
where $P[m]$ is the probability of nuclear spins having the polarization $\sum I_z^{(i)}=m$, $F[m]$ is the degeneracy of this subspace. $\lambda$ enumerates the number of subspaces with the same $j$.
 The probability
for three singlet detections consecutively along $\hat{z}$,$\hat{x}$ and $\hat{z}$ directions in {\it quantum mechanical}(QM) picture is given as
\begin{eqnarray}
\sum_{j,k,m,n,n'} \frac{P[m]}{F[m]}\cos^2[a m\tau_1/2]\cos[an\tau_2/2]\cos[an'\tau_2/2]\nonumber\\ \times\cos^2[ak\tau_3/2] C^{j}_{mn}C^{ j}_{mn'}C^{j}_{k,n}C^{ j}_{k,n'}(F[j]-F[j+1]),\label{eq_inter}
\end{eqnarray}
where $C^{ j}_{m'm}=\langle jm'|\exp[-iJ_y\pi/2|jm\rangle$ is the matrix transforming $J_z$ basis to $J_x$ basis in the subspace specified by the magnitude  $j$ of the angular momentum operator $\sum{\bf I}^{(i)}$ with multiplicity $F[j]-F[j+1]$. 

Equation (\ref{eq_inter}) can be contrasted with the {\it semiclassical}(SC) description of nuclei for which the interference terms are missing, the result of which is
\begin{eqnarray}
\sum_{m,n} P[m]\frac{F[n]}{D}\cos^2[am\tau_1/2]\cos^2[an\tau_2/2]\cos^2[am\tau_3/2], \label{eq_class}
\end{eqnarray}
where $I^{(i)}_z$ is a classical Ising spin taking on values $\{-1/2,1/2\}$ and $D=2^N$ for a spin $1/2$
system.
%where the function $F[m]$ enumerates the number of spin configurations for N classical spins s.t.  $\sum_{i=1}^N I^{(i)}_z=m$ where $I^{(i)}_z$ is a classical spin taking on values $\{-s,-s+1,\ldots,s-1,s\}$.  ${\cal D}=(2s+1)^N$ is the number of possible configurations of nuclear spins.
In the semiclassical case the distribution of nuclear HF field along $\hat{x}$ and $\hat{y}$, is random, whereas it is polarized in $\hat{z}$ direction with distribution $P[m]$ as in the QM case.
%The SC (\ref{eq_class}) and QM distributions (\ref{eq_inter}), reduce to each other in case either $\tau_2=0$ or $\tau_3=0$. For $\tau_1=0$ they reduce to each other only in case there is no polarization of nuclear spins.

\begin{figure}
\psfrag{p}{$P$}
\psfrag{t}{$\sigma \tau_3$}
\includegraphics[width=8.6cm]{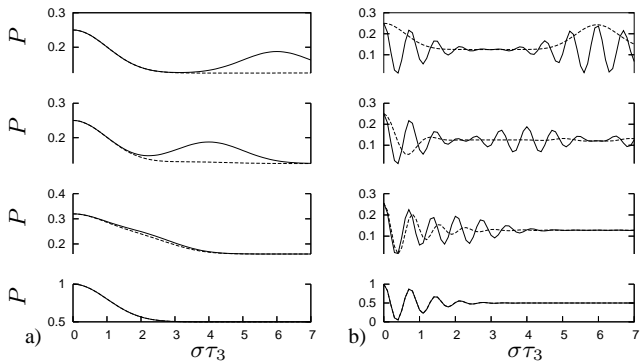}
\caption{Probability of three successive singlet detections along $\hat{z}$,$\hat{x}$,$\hat{z}$ directions respectively, for $N_n=40$ nuclear spins. Initially nuclear spins are polarized along $\hat{z}$ direction with polarizations a) $p=0$ and b)$p=0.8$.  Solid(dashed) curves correspond to semiclassical(quantum mechanical) picture, and from bottom to top HF interaction times $\sigma\tau_1=\sigma\tau_2=0,2,4,6$ in each graph.}
\label{fig_interf}
\end{figure}

In Fig. \ref{fig_interf}, the scheme is exemplified for spin-1/2 nuclei with the number of nuclei $N_n=40$\cite{not1}. Three successive singlet detection probabilities are depicted as a function of $\tau_3$ when the field is along $\hat{z},\hat{x},\hat{z}$ respectively.
In the SC picture (\ref{eq_class}), $\hat{x},\hat{z}$ measurements are independent.
The first and the third measurement results, as a function of $\tau_1$,$\tau_3$ respectively, are maximally correlated and this gives rise to revivals as in (\ref{condprob}).
Whereas in the QM picture (\ref{eq_inter}), these correlations are suppressed due to $\hat{x}$ measurement.
Also in Fig. \ref{fig_interf}b), it is seen that the Overhauser field acting on the electron spin gives rise to Rabi oscillations in the probabilities, which are greatly suppressed in the QM picture.
%At $\tau_3=0$ the probabilities
%coincide, however in the intermediate times they exhibit distinct behavior.
Even in case of unpolarized nuclear spins, SC and QM pictures exhibit distinct behavior. This scheme can also be extended to a a QD with single or a double QD with two electron spins.  
%From equations  (\ref{eq_inter}),(\ref{eq_class}) the recurrence time for the system is $T_{rec}=2\pi N/A$.
%In the Figure-(\ref{fig_interf}) for high polarization states the revivals are also visible, since the evolution is reversible.

Finally, we discuss in brief the feasibility to observe the
predicted phenomena. The duration of the cycle involving electron
spin initialization and measurement is about $10~
\mu$s\cite{Petta05}. Since the nuclear spin coherence time
determined mostly by the nuclear spin diffusion is longer than about
several tens of ms\cite{Paget82}, the bunching for $N$ successive
measurements up to $N>1000$ can be observed. The same holds for the
number of revivals that can be observed. For the demonstration of
quantum interference of nuclear spins changing the
direction of magnetic field before the nuclear spins decohere, may
pose some technical difficulties, especially for a single QD
occupied by two electrons for which a large magnetic field($\sim 10$
T) is needed.

%With a perfect knowledge of the nuclear state, one can totally suppress the
%nuclear decoherence. However, a scheme aimed at exact estimation of the
%nuclear state requires many measurements, during which the system is
%vulnerable to decoherence due to nuclear spin fluctuations.
%Other proposals are on reducing HF induced decoherence via the polarization of
%the nuclear spins and using nuclear spins for quantum information storage\cite{Taylor03a, Imamoglu03,Taylor03b,Stepanenko06,Lai06}.

In summary, we have investigated the dynamics of the electron-nuclei
coupled system in QDs and predicted a couple of new phenomena related
to the correlations induced by the nuclear spins. 
%Especially, the bunching of
%results of the electron spin measurements and the revival in the conditional
%probabilities are salient features. 
The underlying mechanism is the squeezing and
increase in the purity of
the nuclear spin state through the electron spin measurements. This squeezing
is expected to lead to the extension of the electron spin coherence time
because the fluctuation of the nuclear magnetic field due to the
 dipole-dipole interaction would be reduced.
 %By engineering the HF interaction times one can further increase the purity of nuclear spins, thus the number and the strength of revivals.
 Finally we have proposed a scheme for preparing coherent superposition of nuclear spins based on conditional electron spin measurements.
The quantum behavior is manifest even in the case when we have no {\it a priori} knowledge about the initial nuclear spin state.
The predicted results are general and  can be confirmed for electron spins on single and double QDs.

%\bibliographystyle{custom2}
%\bibliography{referansQD}

\begin{thebibliography}{99}
\expandafter\ifx\csname natexlab\endcsname\relax\def\natexlab#1{#1}\fi
\expandafter\ifx\csname bibnamefont\endcsname\relax
  \def\bibnamefont#1{#1}\fi
\expandafter\ifx\csname bibfnamefont\endcsname\relax
  \def\bibfnamefont#1{#1}\fi
\expandafter\ifx\csname citenamefont\endcsname\relax
  \def\citenamefont#1{#1}\fi
\expandafter\ifx\csname url\endcsname\relax
  \def\url#1{\texttt{#1}}\fi
\expandafter\ifx\csname urlprefix\endcsname\relax\def\urlprefix{URL }\fi
\providecommand{\bibinfo}[2]{#2}
\providecommand{\eprint}[2][]{\url{#2}}

\bibitem[{\citenamefont{Loss and DiVincenzo}(1998)\citenamefont{Loss
  et~al.}}]{Loss98}
\bibinfo{author}{\bibfnamefont{D.}~\bibnamefont{Loss}} \bibnamefont{and}
  \bibinfo{author}{\bibfnamefont{D.~P.} \bibnamefont{DiVincenzo}},
  \bibinfo{journal}{Phys. Rev. A} \textbf{\bibinfo{volume}{57}},
  \bibinfo{pages}{120} (\bibinfo{year}{1998}).

\bibitem[{\citenamefont{{A. Imamo\u{g}lu} et~al.}(1999)}]{Imamoglu99}
\bibinfo{author}{\bibnamefont{{A. Imamo\u{g}lu}}}, \bibnamefont{et~al.},
  \bibinfo{journal}{Phys. Rev. Lett.} \textbf{\bibinfo{volume}{83}},
  \bibinfo{pages}{4204} (\bibinfo{year}{1999}).

\bibitem[{\citenamefont{Golovach et~al.}(2004)}]{Golovach04}
\bibinfo{author}{\bibfnamefont{V.~N.} \bibnamefont{Golovach}},
  \bibnamefont{et~al.}, \bibinfo{journal}{Phys. Rev. Lett.}
  \textbf{\bibinfo{volume}{93}}, \bibinfo{pages}{016601}
  (\bibinfo{year}{2004}).

\bibitem[{\citenamefont{Semenov and Kim}(2004)\citenamefont{Semenov
  et~al.}}]{Semenov04}
\bibinfo{author}{\bibfnamefont{Y.~G.} \bibnamefont{Semenov}} \bibnamefont{and}
  \bibinfo{author}{\bibfnamefont{K.~W.} \bibnamefont{Kim}},
  \bibinfo{journal}{Phys. Rev. Lett.} \textbf{\bibinfo{volume}{92}},
  \bibinfo{pages}{026601} (\bibinfo{year}{2004}).

\bibitem[{\citenamefont{Petta et~al.}(2005)}]{Petta05}
\bibinfo{author}{\bibfnamefont{J.~R.} \bibnamefont{Petta}},
  \bibnamefont{et~al.}, \bibinfo{journal}{Science}
  \textbf{\bibinfo{volume}{309}}, \bibinfo{pages}{2180} (\bibinfo{year}{2005}).

\bibitem[{\citenamefont{Koppens et~al.}(2006)}]{Koppens06}
\bibinfo{author}{\bibfnamefont{F.~H.~L.} \bibnamefont{Koppens}},
  \bibnamefont{et~al.}, \bibinfo{journal}{Nature}
  \textbf{\bibinfo{volume}{442}}, \bibinfo{pages}{766} (\bibinfo{year}{2006}).

\bibitem[{\citenamefont{Abragam}(1961)}]{Abragam96}
\bibinfo{author}{\bibfnamefont{A.}~\bibnamefont{Abragam}},
  \emph{\bibinfo{title}{Principles of Nuclear Magnetism}}
  (\bibinfo{publisher}{Oxford U.P.}, \bibinfo{address}{Oxford},
  \bibinfo{year}{1961}).

\bibitem[{\citenamefont{Merkulov et~al.}(2002)}]{Merkulov02}
\bibinfo{author}{\bibfnamefont{I.~A.} \bibnamefont{Merkulov}},
  \bibnamefont{et~al.}, \bibinfo{journal}{Phys. Rev. B}
  \textbf{\bibinfo{volume}{65}}, \bibinfo{pages}{205309}
  (\bibinfo{year}{2002}).

\bibitem[{\citenamefont{Kroutvar et~al.}(2004)}]{Kroutvar04}
\bibinfo{author}{\bibfnamefont{M.}~\bibnamefont{Kroutvar}},
  \bibnamefont{et~al.}, \bibinfo{journal}{Nature}
  \textbf{\bibinfo{volume}{432}}, \bibinfo{pages}{81} (\bibinfo{year}{2004}).

\bibitem[{\citenamefont{Meunier et~al.}(2006)}]{Meunier06}
\bibinfo{author}{\bibfnamefont{T.}~\bibnamefont{Meunier}},
  \bibnamefont{et~al.}, \bibinfo{journal}{eprint:cond-mat/0603794}
  (\bibinfo{year}{2006}).

\bibitem[{\citenamefont{Taylor et~al.}(2003{\natexlab{a}})}]{Taylor03b}
\bibinfo{author}{\bibfnamefont{J.~M.} \bibnamefont{Taylor}},
  \bibnamefont{et~al.}, \bibinfo{journal}{Phys. Rev. Lett.}
  \textbf{\bibinfo{volume}{91}}, \bibinfo{pages}{246802}
  (\bibinfo{year}{2003}{\natexlab{a}}).

\bibitem[{\citenamefont{Imamoglu et~al.}(2003)}]{Imamoglu03}
\bibinfo{author}{\bibfnamefont{A.}~\bibnamefont{Imamoglu}},
  \bibnamefont{et~al.}, \bibinfo{journal}{Phys. Rev. Lett.}
  \textbf{\bibinfo{volume}{91}}, \bibinfo{pages}{017402}
  (\bibinfo{year}{2003}).

\bibitem[{\citenamefont{Klauser et~al.}(2005)}]{Klauser05}
\bibinfo{author}{\bibfnamefont{D.}~\bibnamefont{Klauser}},
  \bibnamefont{et~al.}, \bibinfo{journal}{Phys. Rev. B}
  \textbf{\bibinfo{volume}{73}}, \bibinfo{pages}{205302}
  (\bibinfo{year}{2005}).

\bibitem[{\citenamefont{Giedke et~al.}(2006)}]{Giedke05}
\bibinfo{author}{\bibfnamefont{G.}~\bibnamefont{Giedke}}, \bibnamefont{et~al.},
  \bibinfo{journal}{Phys. Rev. A} \textbf{\bibinfo{volume}{74}},
  \bibinfo{pages}{032316} (\bibinfo{year}{2006}).

\bibitem[{\citenamefont{Stepanenko et~al.}(2006)}]{Stepanenko06}
\bibinfo{author}{\bibfnamefont{D.}~\bibnamefont{Stepanenko}},
  \bibnamefont{et~al.}, \bibinfo{journal}{Phys. Rev. Lett.}
  \textbf{\bibinfo{volume}{96}}, \bibinfo{pages}{136401}
  (\bibinfo{year}{2006}).

\bibitem[{\citenamefont{Rudner and Levitov}(2006)\citenamefont{Rudner
  et~al.}}]{Rudner06}
\bibinfo{author}{\bibfnamefont{M.~S.} \bibnamefont{Rudner}} \bibnamefont{and}
  \bibinfo{author}{\bibfnamefont{L.~S.} \bibnamefont{Levitov}},
  \bibinfo{journal}{cond-mat/0609409}  (\bibinfo{year}{2006}).

\bibitem[{\citenamefont{Taylor et~al.}(2003{\natexlab{b}})}]{Taylor03a}
\bibinfo{author}{\bibfnamefont{J.~M.} \bibnamefont{Taylor}},
  \bibnamefont{et~al.}, \bibinfo{journal}{Phys. Rev. Lett.}
  \textbf{\bibinfo{volume}{90}}, \bibinfo{pages}{206803}
  (\bibinfo{year}{2003}{\natexlab{b}}).

\bibitem[{\citenamefont{Coish and Loss}(2005)\citenamefont{Coish
  et~al.}}]{Coish05}
\bibinfo{author}{\bibfnamefont{W.~A.} \bibnamefont{Coish}} \bibnamefont{and}
  \bibinfo{author}{\bibfnamefont{D.}~\bibnamefont{Loss}},
  \bibinfo{journal}{Phys. Rev. B} \textbf{\bibinfo{volume}{72}},
  \bibinfo{pages}{125337} (\bibinfo{year}{2005}).

\bibitem[{\citenamefont{Cakir and Takagahara}(2006)\citenamefont{Cakir
  et~al.}}]{Cakir06}
\bibinfo{author}{\bibfnamefont{O.}~\bibnamefont{Cakir}} \bibnamefont{and}
  \bibinfo{author}{\bibfnamefont{T.}~\bibnamefont{Takagahara}},
  \bibinfo{journal}{cond-mat/0609217}  (\bibinfo{year}{2006}).

\bibitem[{\citenamefont{Hanson et~al.}(2003)}]{Hanson03}
\bibinfo{author}{\bibfnamefont{R.}~\bibnamefont{Hanson}}, \bibnamefont{et~al.},
  \bibinfo{journal}{Phys. Rev. Lett.} \textbf{\bibinfo{volume}{91}},
  \bibinfo{pages}{196802} (\bibinfo{year}{2003}).

\bibitem[{\citenamefont{Dutt et~al.}(2005)}]{Dutt05}
\bibinfo{author}{\bibfnamefont{M.~V.~G.} \bibnamefont{Dutt}},
  \bibnamefont{et~al.}, \bibinfo{journal}{Phys. Rev. Lett.}
  \textbf{\bibinfo{volume}{94}}, \bibinfo{pages}{227403}
  (\bibinfo{year}{2005}).

\bibitem[{\citenamefont{Atat{\"u}re et~al.}(2006)}]{Atature06}
\bibinfo{author}{\bibfnamefont{M.}~\bibnamefont{Atat{\"u}re}},
  \bibnamefont{et~al.}, \bibinfo{journal}{Science}
  \textbf{\bibinfo{volume}{312}}, \bibinfo{pages}{551} (\bibinfo{year}{2006}).

\bibitem[{\citenamefont{Fujisawa et~al.}(2002)}]{Fujisawa02}
\bibinfo{author}{\bibfnamefont{T.}~\bibnamefont{Fujisawa}},
  \bibnamefont{et~al.}, \bibinfo{journal}{Nature}
  \textbf{\bibinfo{volume}{419}}, \bibinfo{pages}{278} (\bibinfo{year}{2002}).

\bibitem[{\citenamefont{Hanson et~al.}(2005)}]{Hanson05}
\bibinfo{author}{\bibfnamefont{R.}~\bibnamefont{Hanson}}, \bibnamefont{et~al.},
  \bibinfo{journal}{Phys. Rev. Lett.} \textbf{\bibinfo{volume}{94}},
  \bibinfo{pages}{196802} (\bibinfo{year}{2005}).

\bibitem[{\citenamefont{{T. Meunier et. al.}}(2006)}]{Meunier06b}
\bibinfo{author}{\bibnamefont{{T. Meunier et. al.}}} (\bibinfo{year}{2006}),
  \bibinfo{note}{e-print: cond-mat/0609726}.

\bibitem[{not()}]{not1}
\bibinfo{note}{Qualitative features are not dependent on $N_n$ when $1\gg
  1/\sqrt{N_n}$. Results will be published elsewhere.}

\bibitem[{\citenamefont{Paget}(1982)}]{Paget82}
\bibinfo{author}{\bibfnamefont{D.}~\bibnamefont{Paget}},
  \bibinfo{journal}{Phys. Rev. B} \textbf{\bibinfo{volume}{25}},
  \bibinfo{pages}{4444} (\bibinfo{year}{1982}).
\end{thebibliography}

\end{document}